\documentclass[letterpaper, 10 pt, conference]{ieeeconf}  
\IEEEoverridecommandlockouts
\renewcommand{\baselinestretch}{0.97}

\usepackage{graphicx}
\usepackage{lipsum}
\usepackage{amsmath} 
\usepackage{amssymb} 
\usepackage{subcaption}
\usepackage{comment}
\usepackage{url}

\usepackage{fancyhdr}

\fancypagestyle{firstpage}{%
    \fancyhf{}%
    \fancyheadoffset[L,R]{0.8cm}%
    \fancyhead[C]{%
        \parbox{\headwidth}{%
            \vspace{-35pt}
            \scriptsize
            \textcopyright{} 2026 IEEE. Personal use of this material is permitted. Permission from IEEE must be obtained for all other uses, in any current or future media, including reprinting/republishing this material for advertising or promotional purposes, creating new collective works, for resale or redistribution to servers or lists, or reuse of any copyrighted component of this work in other works.
            
            \smallskip
            
            \noindent
            This is the author's accepted manuscript of the paper: 

            \noindent
            D. Costa, F. Cerrito, M. Canale and C. Novara, "Unifying Decision-Making and Trajectory-Planning in Unsignalized Intersections Using Time-Varying Potential Fields*," 2026 34th Mediterranean Conference on Control and Automation (MED), Ancona, Italy, 2026, pp. 251-256, doi: 10.1109/MED70602.2026.11598041.
            
            \noindent
            The final published version is available in IEEE Xplore.
        }%
    }%
}

\title{\LARGE \bf
Unifying Decision-Making and Trajectory-Planning in Unsignalized Intersections Using Time-Varying Potential Fields$^\star$
}

\author{David Costa$^\dag$, Francesco Cerrito$^\ddag$, Massimo Canale$^\dag$ and Carlo Novara$^\dag$%
\thanks{$^\star$ This work is part of the project Piano Nazionale di Ripresa e Resilienza (PNRR)- Next Generation Europe, which has received funding from the Italian Ministry of University and Research – DM 117/2023.}
\thanks{$^\dag$Dipartimento di Elettronica e Telecomunicazioni, Politecnico di Torino, 10129, Turin, Italy
        {\tt\small david.costa@polito.it}, {\tt\small massimo.canale@polito.it}, {\tt\small carlo.novara@polito.it}}%
\thanks{$^\ddag$Dipartimento di Automatica e Informatica, Politecnico di Torino, 10129, Turin, Italy
        {\tt\small francesco.cerrito@polito.it}}%
}

\begin{document}
\maketitle
\thispagestyle{firstpage}
\pagestyle{empty}

\begin{abstract}
This paper presents a novel framework for integrated Decision-Making (DM) and Trajectory Planning (TP) for automated vehicles at unsignalized intersections. The approach leverages a Finite Horizon Optimal Control Problem (FHOCP) that employs Time-Varying Artificial Potential Fields (TV-APF). By utilizing short-horizon motion prediction and a dedicated conflict-zone occupancy coefficient, the framework suitably accounts for potential collisions within the FHOCP. The proposed method effectively unifies DM and TP, ensuring the generation of a feasible and safe reference trajectory. Simulation results in multi-vehicle traffic scenarios demonstrate the effectiveness of the approach.
\end{abstract}

\section{Introduction}
Unsignalized intersections are among the most demanding scenarios for autonomous vehicles~\cite{AlSharman2024,ChenHu2024}.
In the absence of traffic signals, vehicles must rely on local priority rules and real-time observations of surrounding traffic to decide when and how to proceed through the conflict zone.
Multiple vehicles with unknown intentions may approach the intersection simultaneously from different directions, and the variety of possible entry--exit branch combinations creates a combinatorial space of interactions that challenges both safety and efficiency.
These factors place a serious demand on the Trajectory Planning (TP) and Decision-Making (DM) capability of autonomous driving systems. Existing solutions can be broadly categorized into two groups: infrastructure-based coordination methods and ego-centric approaches.

In centralized coordination or infrastructure-based methods, a roadside controller or intersection manager assigns crossing orders and time slots to approaching vehicles~\cite{RiosTorresMalikopoulos2017,ZhaoMACPO2024}.
Reservation-based schemes and cooperative scheduling algorithms can, in principle, guarantee collision-free traversal and maximize intersection throughput.
However, these approaches often require full vehicle-to-infrastructure communication and assume that all vehicles are connected and compliant, limiting their applicability in mixed-traffic environments where human-driven vehicles coexist with autonomous ones, or when the infrastructure is degraded or unavailable.

On the other hand, a range of ego-based planning methods has also been investigated.
Game-theoretic formulations model strategic interactions among multiple vehicles, capturing the coupled decision processes inherent to unsignalized intersections~\cite{TianGameTheory2022}, but they often require computationally expensive equilibrium calculations that are challenging for real time usage.
Reinforcement or deep learning methods can, in principle, learn flexible intersection navigation policies directly from simulation or data~\cite{LiPeng2024,ChenXu2024,TangUncertainty2023}, yet they typically require extensive training, may lack formal safety guarantees, and remain predominantly focused on high-level behavioral decisions rather than integrated TP~\cite{AlSharman2024}.
Optimization-based approaches (c.f. \cite{MPCsw}), including Model Predictive Control (MPC) and Finite-Horizon Optimal Control Problem (FHOCP), explicitly encode vehicle dynamics, safety constraints, and comfort objectives, making them attractive for real-time TP in interactive driving scenarios.
In this setting, Artificial Potential Fields (APFs), see e.g. \cite{APF0}, in which obstacles are modeled using repulsive fields and goals using attractive fields, offer a computationally lightweight solution, as they can be embedded directly into a planning objective.
Recent work has incorporated APFs as reward-shaping signals within sampling-based MPC at unsignalized intersections~\cite{Zhang2025DAWM}, but in such approaches the APF is used as an auxiliary component rather than the core DM and TP mechanism.

Across such ego-based approaches, DM and TP are usually realized through staged or hierarchical architectures.
In such a context, a high-level module implemented, for example, as a finite-state machine, a rule-based logic, or a learned policy, typically selects a behavioral mode or maneuver class. Then, a downstream planner or controller generates a feasible trajectory consistent with that choice~\cite{JeongYi2021,Raja2024}.
Although this modular decomposition simplifies the overall system design, it may introduce potential inconsistencies, since transition between discrete modes may result in infeasibility for the planner under the current dynamic constraints, and rigid mode boundaries can prevent behavior adaptation as the traffic situation evolves.
These issues are especially pronounced at unsignalized intersections, where the appropriate behavior depends on the continuously changing positions and velocities of multiple interacting vehicles and cannot always be captured by a finite set of predefined modes.

Although each of the approaches described above addresses important aspects of the intersection navigation problem, the integration of DM and TP into a single optimization problem, without relying on discrete behavioral switching, remains largely an open challenge~\cite{AlSharman2024}.

This paper proposes a unified framework in which both DM and TP are obtained by solving a suitable FHOCP. The method builds on the Time-Varying Artificial Potential Field (TV-APF) previously introduced in~\cite{CostaTVAPF_prev} for highway and country-road scenarios and extends it to unsignalized intersections. In particular, in \cite{CostaTVAPF_prev}, the FHOCP is the core part of a hierarchical control architecture that, based on the prediction of the behavior of the different traffic actors described by TV-APFs, decides which maneuver is the most appropriate to perform a specific driving task, e.g., to cross an intersection. This is obtained through the computation of a safe collision free trajectory and the corresponding control action in terms of acceleration and steer, see \cite{CostaTVAPF_prev} for details. 
Specifically, our contribution consists in defining an intersection-oriented TV-APF whose amplitude is modulated by a predicted conflict-zone occupancy signal obtained from short-horizon motion predictions of surrounding vehicles. The resulting potential field is continuously differentiable and is incorporated directly into the cost function of the FHOCP. As a result, behaviors such as yielding, stopping, decelerating, or crossing the intersection naturally emerge from a single optimization problem, without requiring discrete decision logic or mode switching. \\
Simulation tests are performed using the hierarchical control architecture in \cite{CostaTVAPF_prev}, where a global planner provides a reference path and intersection metadata, a local planner solves the TV-APF-based FHOCP for each replanning step, and a motion controller tracks the resulting trajectory.
\subsubsection*{Notation}
The set of integers over a specific interval $[a, b]$ is denoted as: $\mathbb{N}_{[a,b]} = \{i \in \mathbb{N} \mid a \leq i \leq b\}$.\\
The standard FHOCP notation is used to denote predictions. The predicted value $x$ at time instant $k+\ell$, computed using information available at the current time instant $k$, is denoted by $x(k+\ell|k)$.
\vspace{-5px}
\section{Problem Formulation}
We focus on the DM and TP problem for an automated vehicle approaching an unsignalized intersection. The intersection topology and the tactical decision of the AV, e.g., the reference path associated with the navigation goal, are assumed to be known a priori. The core problem is to compute an ego reference trajectory that ensures safe maneuvering through the conflict zone. In practical terms, the planner must decide whether to stop and wait for other vehicles to pass, or whether the conflict zone will be free in time to allow a safe crossing.

At unsignalized intersections, right-of-way (and thus DM) is not enforced by an external infrastructure (e.g., traffic lights or a centralized coordination). Instead, vehicles rely on local priority rules (e.g., yield-to-the-right) and on real-time observations of other road users. Multiple surrounding vehicles (hereafter named \emph{actors}) with partially unknown intentions may approach the same conflict zone, and different combinations of entry and exit branches yield many possible crossing configurations. This combinatorial growth makes purely rule-based DM difficult to scale while maintaining both safety and efficiency. 
Incorporating DM \emph{within a suitable optimization problem} mitigates both issues by allowing the solver itself to trade off among the different behavioral (and thus behavior-related trajectories) alternatives in a single shot. In this work, we employ PFs within an optimization-based trajectory planner to obtain a unified DM/TP formulation for unsignalized intersections. Building on the TV-APF framework introduced in~\cite{CostaTVAPF_prev}, we define an intersection-specific TV-APF whose amplitude is modulated by a predicted conflict-zone occupancy signal. The resulting potential can be included directly in the objective function of the planner's FHOCP.
\vspace{-3px}
\section{Preliminaries}\label{sec:prel}
\vspace{-0.05cm}
Since this paper builds upon the unified DM and TP framework introduced in our previous work~\cite{CostaTVAPF_prev}, its main elements are briefly summarized here to keep the paper self-contained. We refer to \cite{CostaTVAPF_prev} for a detailed description of such a framework.

The DM/TP problem is formulated as a FHOCP that is solved periodically every $T_P$ seconds. In this context, APF and TV-APF are used over a prediction horizon of $N_P$ steps to generate a safe trajectory corresponding to the most appropriate maneuver to be performed, e.g., lane keeping, yielding/following, overtaking. To construct the FHOCP cost for a generic planning step $k$, the following information is assumed to be available:
\begin{itemize}
    \item a global smooth reference path $\Pi\in\mathcal{C}^2$ connecting the centerline of the roads used by the ego vehicle to reach its goal;
    \item road attributes (e.g., number of lanes, lane width, and speed limits);
    \item the matrix $\Xi$, containing the ordered predictions of all $n_o$ surrounding actors within a suitable perception range. For a generic actor $j\in\mathbb{N}_{[1,n_o]}$, its prediction is denoted by $\Xi_j(k+\ell|k)$ and contains the estimated position $p_j(k+\ell|k)$ and velocity $v_j(k+\ell|k)$ at each step $\ell\in \mathbb{N}_{[0,N_P-1]}$.
\end{itemize}
Two reference frames are adopted to simplify the formulation and solution of the FHOCP:
\begin{itemize}
    \item \emph{Global Cartesian frame}: used to represent the environment and road geometry in $(X,Y)$ coordinates.
    \item \emph{Local Frenet frame}: defined with respect to the reference path $\Pi$, where the vehicle position is represented by the curvilinear abscissa $s$ and the lateral offset $d$. This representation simplifies the computation of distances from lanes and actors.
\end{itemize}
Vehicle motion over the horizon is predicted through a standard kinematic model as in \cite{CostaTVAPF_prev}, that can be described by discrete-time dynamic model
\begin{equation}
\chi(k+1)=f(\chi(k),u(k)),
\label{eq:vh_dyn_d}
\end{equation}
where $\chi$ represents the vehicle state, i.e., position, orientation, velocity, and $u$ is the control input vector, i.e., longitudinal acceleration and steering rate. The optimization problem is therefore solved with respect to the predicted state sequence $\chi$ and input sequence $U$. The optimized ego state sequence is obtained as
\begin{equation}
\chi^*=\arg\min_{U,\chi}\mathcal{J}(U,\chi,\Xi)
\label{eq:nlp_complete}
\end{equation}
subject to the vehicle dynamics~\eqref{eq:vh_dyn_d}, admissible state, input, and input-rate bounds, safety constraints, and terminal conditions. The cost $\mathcal{J}$ in \eqref{eq:nlp_complete} is defined as
\begin{align}
\mathcal{J}(U,\chi,\Xi) = \sum_{\ell=0}^{N_P-1} &\left(K_vW_v + K_bW_b + \right.\label{eq:cost}\\
\nonumber &\left. K_lW_l + K_cW_c + K_oO\right),\\
O(\chi,\Xi,\ell) =&\ \sum_{j=1}^{n_o} W_{o,j}\!\left(\chi(k+\ell|k),\Xi_j(k+\ell|k)\right).
\label{eq:TV-APF1}
\end{align}
where each weighted objective term penalizes undesirable states and control actions. In particular:
\begin{itemize}
    \item $W_v(\chi)$, weighted by $K_v$, penalizes deviations from the desired speed profile;
    \item $W_b(\chi)$, weighted by $K_b$, enforces lane boundary constraints;
    \item $W_l(\chi)$, weighted by $K_l$, promotes occupation of the rightmost free lane;
    \item $W_c(\chi,U)$, weighted by $K_c$, penalizes high lateral acceleration;
    \item $O(\chi,\Xi,\ell)$, weighted by $K_o$, accounts for obstacle avoidance; $W_{o,j}$ is the TV-APF contribution induced by actor $j$ and is evaluated from the predicted ego state and the predicted actor state at step $\ell$.
\end{itemize}

Unlike hierarchical approaches, where DM and TP are treated as separate layers, the proposed framework integrates both processes within a single optimization problem.

\begin{figure}
    \centering
    \includegraphics[width=0.45\linewidth]{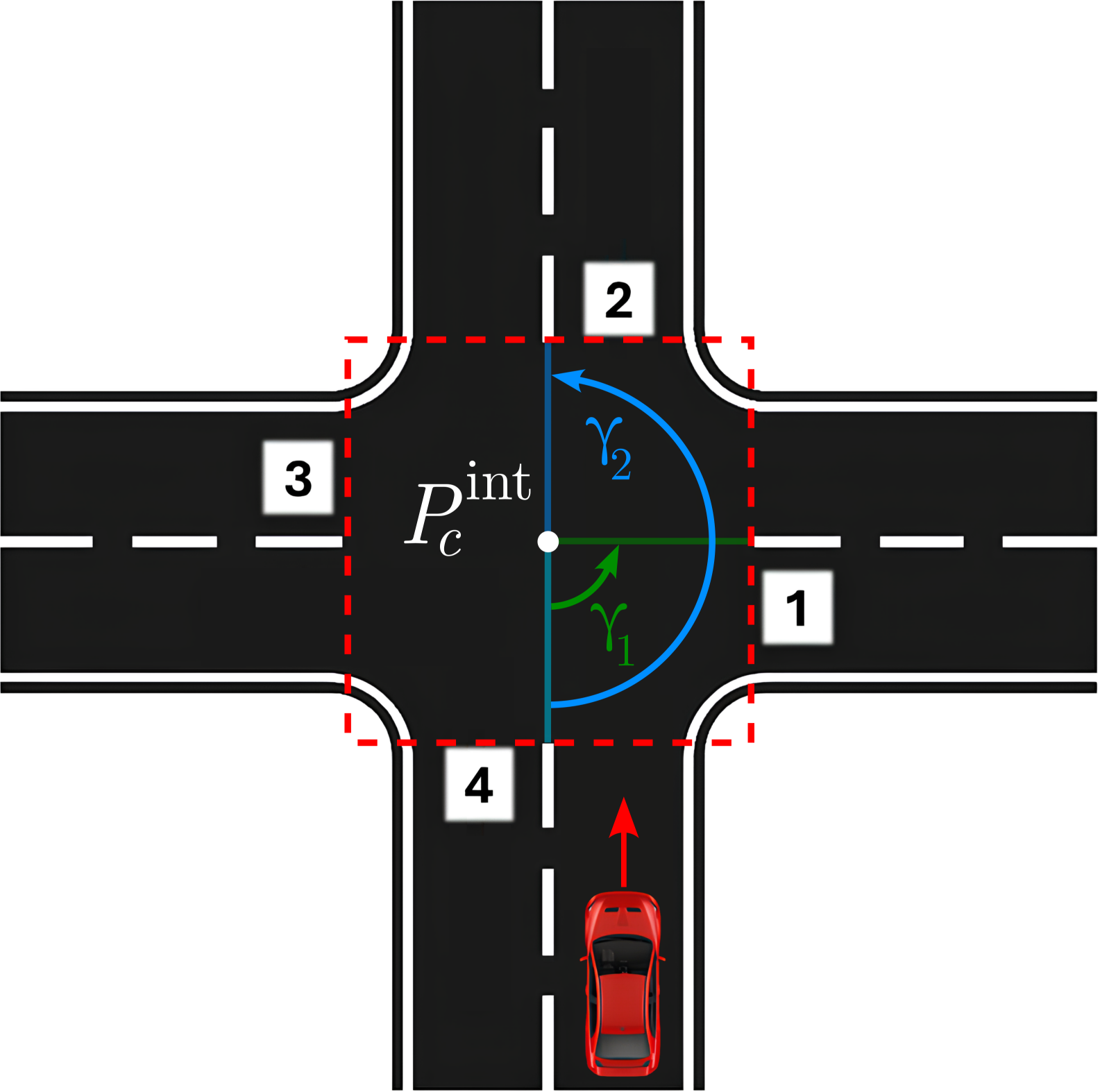}
    \caption{Four-way intersection showing the ego vehicle, the intersection center $P_c^{int}$, the road branches and corresponding angles $\gamma_1$ (in green) and $\gamma_2$ (in blue). The conflict zone is highlighted in red.}
    \label{fig:intersection-description}
\end{figure}

Within this framework, the objective of this paper is to properly shape a dedicated TV-APF to efficiently handle intersection interactions. The proposed approach does not modify the overall planner structure; instead, it extends the existing formulation by introducing an additional term in~\eqref{eq:cost} to account for intersection management within the same optimization problem.\\
To effectively handle intersections, the FHOCP needs some additional a-priori information describing the road crossing. In particular, we assume the following data to be available: (a) the position of the intersection center in global coordinates, $P_c^{\mathrm{int}}=[x_c^{\mathrm{int}},y_c^{\mathrm{int}}]^\top$ (Fig.~\ref{fig:intersection-description}), together with its corresponding longitudinal Frenet coordinate $s_{\mathrm{int}}$ along the ego reference path $\Pi$; (b) a description of the intersection branches; and (c) the selected exit branch for the ego vehicle.

\section{Intersection-Aware Actor Selection and Occupancy Estimation}
\label{sec:int_actor_selection}
To define and activate the intersection TV-APF consistently, two prior steps are required: (i) selecting only the traffic actors that can influence the ego vehicle behavior at the intersection and (ii) predicting their motion in a neighborhood of the conflict zone. 
\vspace{-2px}
\subsection{Selection of interacting actors at an intersection}
\label{sec:int_selection}
Only a subset of the surrounding vehicles is relevant for the intersection interaction: (i) vehicles that have priority over the ego vehicle (the ego must yield), and (ii) vehicles that do not have priority but may still intersect the ego vehicle's intended path and thus become dangerous if they fail to stop. All other actors are ignored to reduce computational load and to avoid overly conservative behavior.

\begin{figure}[]
    \centering
    \includegraphics[width=0.32\linewidth]{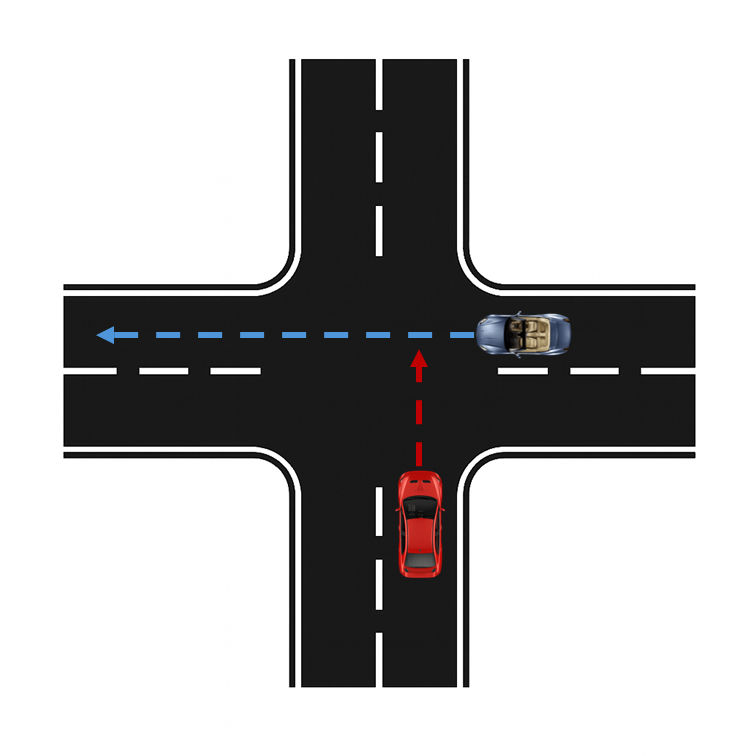}
    \hfill
    \includegraphics[width=0.32\linewidth]{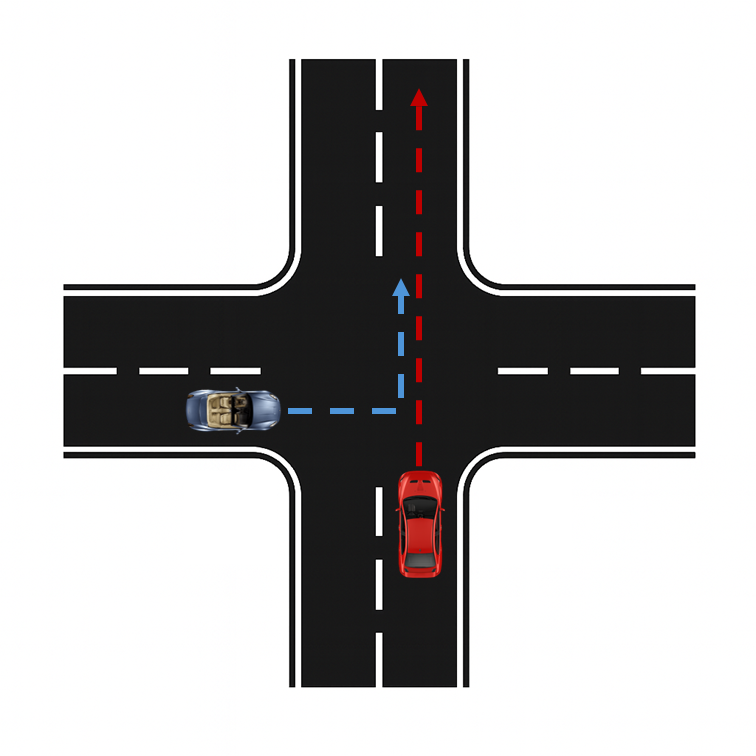}
    \hfill
    \includegraphics[width=0.32\linewidth]{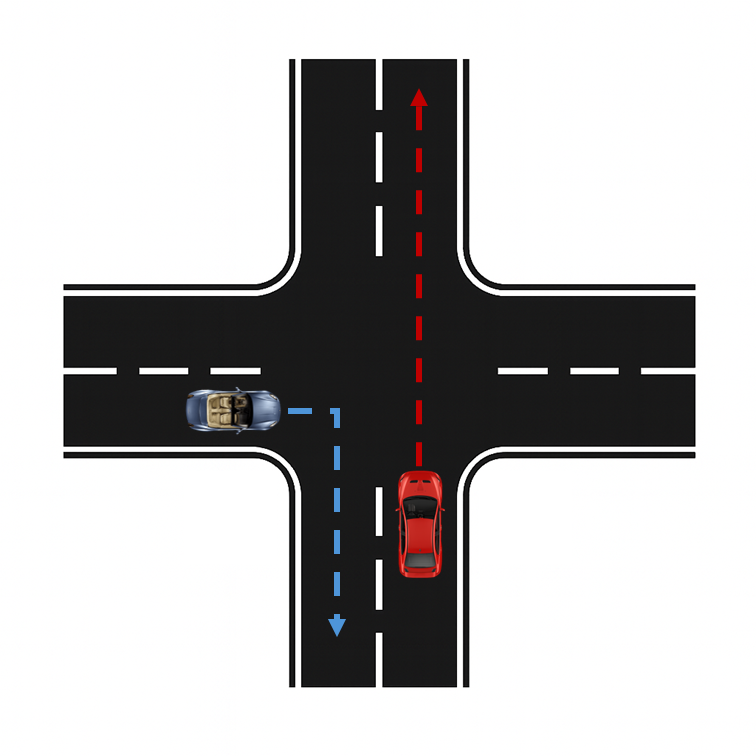}
    \caption{Two-vehicle interaction patterns at an unsignalized intersection (ego vehicle in red, actor in blue): (left) ego vehicle must yield to the right; (center) ego vehicle has priority, but trajectories intersect; (right) ego vehicle has priority and trajectories do not interfere.}
    \label{fig:row_cases}
\end{figure}

\begin{figure}[]
    \centering
    \includegraphics[width=0.4\linewidth]{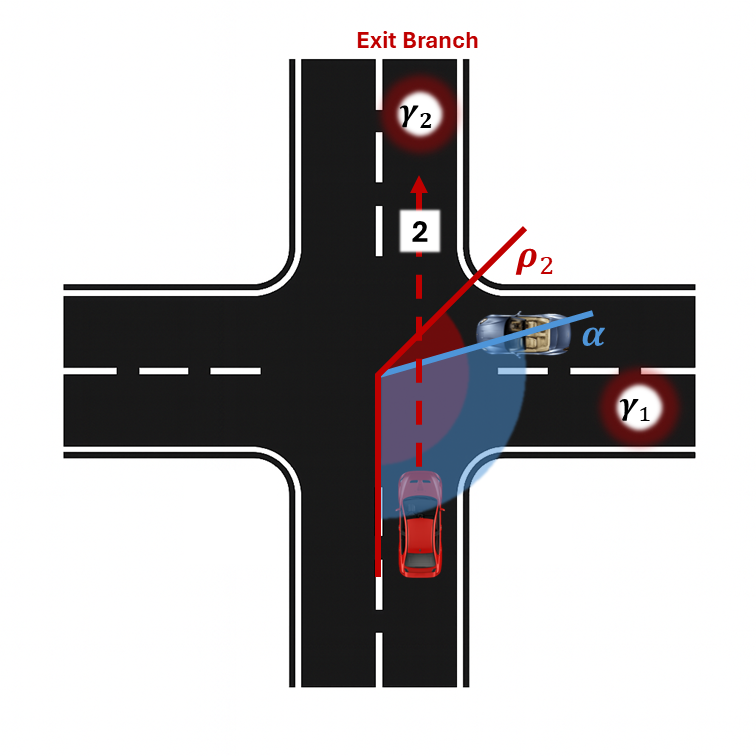}
    \hfill
    \includegraphics[width=0.4\linewidth]{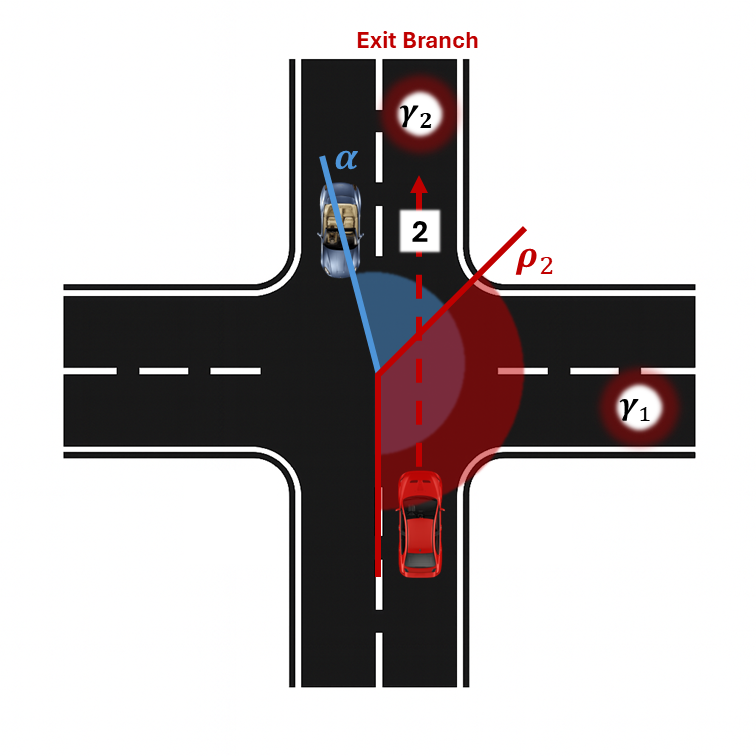}
    \caption{Right-of-way selection using the angular threshold \(\rho_e\) with $e=2$: actors (blue) within the right-side sector are selected; actors outside the sector are ignored for right-of-way purposes.}
    \label{fig:angle_right}
\end{figure}

\begin{figure}[]
    \centering
    \includegraphics[width=0.4\linewidth]{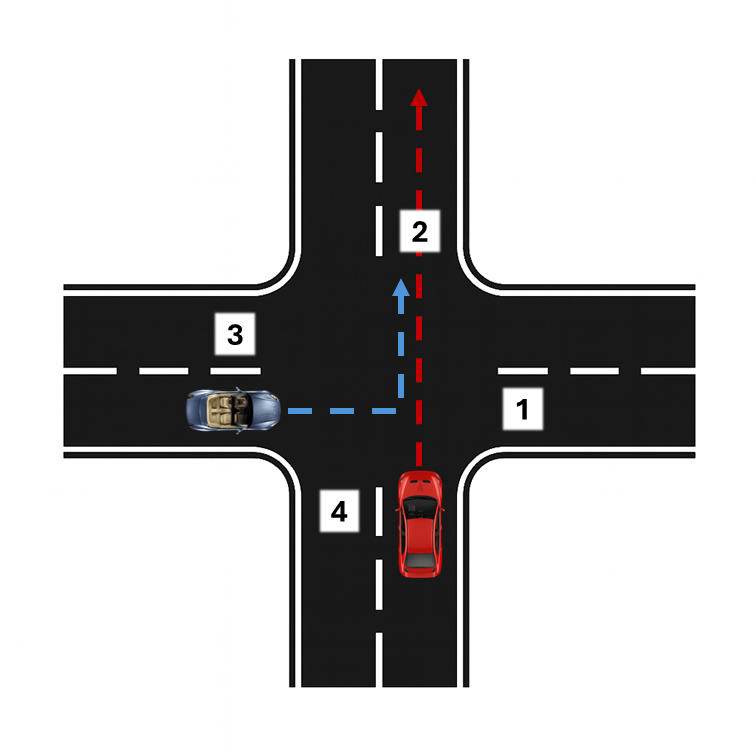}
    \hfill
    \includegraphics[width=0.4\linewidth]{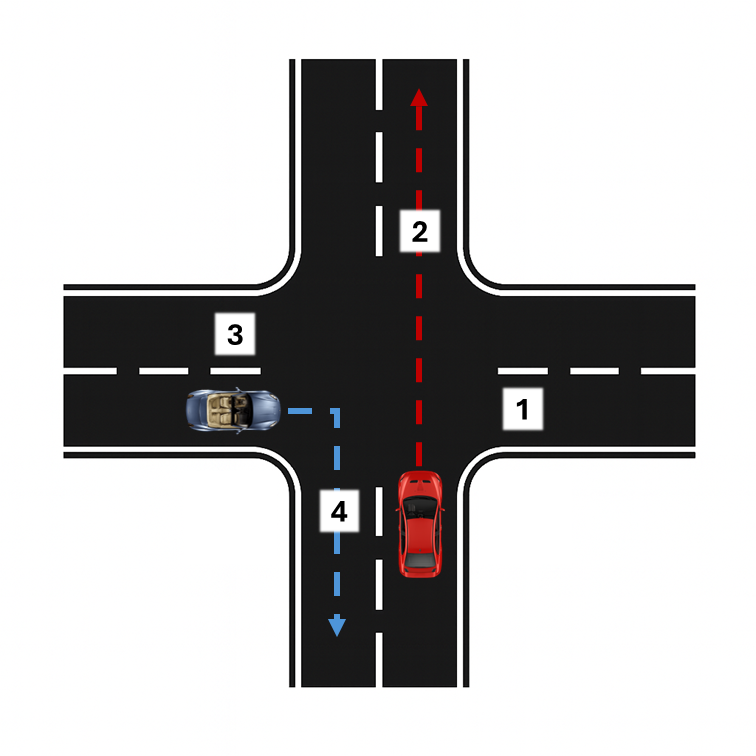}
    \caption{Potentially interfering actor selection using the intended exit branch index: actors (blue) with \(b^{\mathrm{out}}_j \le b^{\mathrm{out}}_{\mathrm{ego}}\) may cross/overlap the ego vehicle path (red) and are selected; higher-index exits are ignored for this criterion.}
    \label{fig:exit_branch}
\end{figure}

Let an intersection be composed of \(n_b\) road branches and assume, without loss of generality, that turning restrictions are not enforced (i.e., all exits are admissible). Starting from the ego vehicle incoming direction and moving counterclockwise, the available exit branches are ordered and assigned increasing indices ($i \in \mathbb{N}_{[1,n_b]}$) (Fig. \ref{fig:intersection-description}). The same indexing convention is used to interpret the intended exits of surrounding vehicles. 
We associate each branch \(i\) with an orientation angle \(\gamma_i\in[0,2\pi)\), measured counterclockwise around the intersection center as shown in Fig. \ref{fig:intersection-description}. 
Then, for each surrounding vehicle \(j\), we assume that its intended exit branch can either be inferred from turn-signal detection or communicated by V2X/V2V, and indexed as \(b^{\mathrm{out}}_j\). Instead, the intended exit of the ego vehicle is denoted by \(b^{\mathrm{out}}_{\mathrm{ego}}\). Although intersections admit many combinations of incoming/outgoing branches, two-vehicle configurations fall into three practical categories (Fig. \ref{fig:row_cases}):
\begin{enumerate}
    \item the ego vehicle must yield to an actor approaching from its right-hand side;
    \item the ego vehicle has priority, but the other actor trajectory intersects its intended path (risk if the other actor fails to stop);
    \item the ego vehicle has priority, and the other actor trajectory does not interfere with it.
\end{enumerate}
The selection criterion is to target the first two categories only.
For compactness, we describe the selection through actor sets. Let \(\mathcal{A}\) be the set of perceived vehicles for the intersection considered. The selected set \(\mathcal{A}_{\mathrm{sel}}\) is built as
\begin{equation}
    \mathcal{A}_{\mathrm{sel}} = \mathcal{A}_{\mathrm{ROW}} \cup \mathcal{A}_{\mathrm{X}},
    \label{eq:actor_sets}
\end{equation}
where \(\mathcal{A}_{\mathrm{ROW}}\) contains right-of-way actors and \(\mathcal{A}_{\mathrm{X}}\) contains potentially interfering actors. Such sets are defined in the following.
\paragraph{Right-of-way actors (yield-to-the-right)}
Given the exit branch of the ego vehicle \(e = b^{\mathrm{out}}_{\mathrm{ego}}\), we define a right-of-way angular threshold
\begin{equation}
    \rho_e = \frac{\gamma_{e} + \gamma_{e-1}}{2}\, \text{.}
    \label{eq:rho_e}
\end{equation}
Let \(\alpha_j\) be the polar angle of vehicle \(j\) measured around the intersection center (Fig. \ref{fig:angle_right}). Vehicles with \(\alpha_j < \rho_e\) are classified as right-of-way vehicles, meaning that the ego should yield to them for the considered maneuver. In practice, equality cases \(\alpha_j \approx \rho_e\) can be treated conservatively as yielding conditions, and angles should be compared after a consistent wrap to \([0,2\pi)\). More explicitly, the actor angle is computed as
\begin{equation}
    \alpha_j = \mathrm{wrap}_{2\pi}\!\Big(\mathrm{atan2}(y_j-y_c^{\mathrm{int}},\,x_j-x_c^{\mathrm{int}})\Big)\,\text{.}
    \label{eq:alpha_j}
\end{equation}
where \(\mathrm{wrap}_{2\pi}(\cdot)\) maps angles to \([0,2\pi)\). The right-of-way actor set is then
\begin{equation}
    \mathcal{A}_{\mathrm{ROW}} = \big\{ j \in \mathcal{A} \,:\, \alpha_j \le \rho_e \big\}.
    \label{eq:actor_set_row}
\end{equation}
\vspace{-10px}
\paragraph{Potentially interfering vehicles (path-crossing risk)}
Vehicles not classified as right-of-way actors may still intersect the ego vehicle path depending on their intended exit (Fig. \ref{fig:exit_branch}). We mark vehicle \(j\) as a potential interferer if
\begin{equation}
    b^{\mathrm{out}}_j \le b^{\mathrm{out}}_{\mathrm{ego}}\,\text{,}
    \label{eq:interferer_rule}
\end{equation}
that is, the case in which another actor may cross the ego vehicle trajectory, even though the ego vehicle has the right of way. To avoid considering already-resolved conflicts, stationary vehicles that have yielded to the ego vehicle and are waiting at the stop line (e.g. \(\|v_j\|<v_{\varepsilon}\)) are excluded from the set of \emph{potentially interfering} vehicles. We can then define the set $\mathcal{A}_{\mathrm{X}}$ as
\begin{equation}
    \mathcal{A}_{\mathrm{X}} =
    \Big\{ j \in \mathcal{A} \,:\,
    \alpha_j > \rho_e,\;
    b^{\mathrm{out}}_j \le b^{\mathrm{out}}_{\mathrm{ego}},\;
    \|v_j\| \ge v_{\varepsilon}
    \Big\}\,\text{.}
    \label{eq:actor_set_x}
\end{equation}
\vspace{-10px}
\subsection{Short-horizon prediction and signed distance to the conflict zone}
\label{sec:int_prediction}
Based on the current speed $v_j$ of the selected actors $\mathcal{A}_{\mathrm{sel}}$, their positions are predicted over a short horizon to assess whether they will occupy the intersection area concurrently with the ego vehicle. 
In this context, we approximate the actor distance to the center of the intersection using the Euclidean distance. Assuming a constant-velocity model over the horizon \(H_P = N_P T_s\), the predicted position of actor \(j\) in global coordinates $p_j=(x_j,y_j)^\top$ at step \(k+1\) is
\begin{equation}
    p_j(k+1) = p_j(k) + T_s\, v_j\, \text{,}\qquad k \in \mathbb{N}_{[1,N_P-1]}\,\text{.}
    \label{eq:cv_prediction}
\end{equation}
We define a signed distance (negative while approaching, positive while leaving) for each actor $j$ from the intersection center \(P_c^{\mathrm{int}}\) as:
\begin{equation}
    d_{\mathrm{int},j}(k) =
    \mathrm{sign}\!\Big(v_j^\top \big(p_j(0) - P_c^{\mathrm{int}}\big)\Big)\,
    \big\| p_j(k) - P_c^{\mathrm{int}} \big\|_2 \text{.}
    \label{eq:signed_distance}
\end{equation}
Since predictions are updated at each re-planning cycle $T_P$, such a  simple model is sufficient to capture the principal effect needed for occupancy estimation, reducing the computational burden.
\subsection{Intersection occupancy coefficient}
\label{sec:int_occupancy}
\begin{figure}[]
    \centering
    \includegraphics[width=0.7\linewidth]{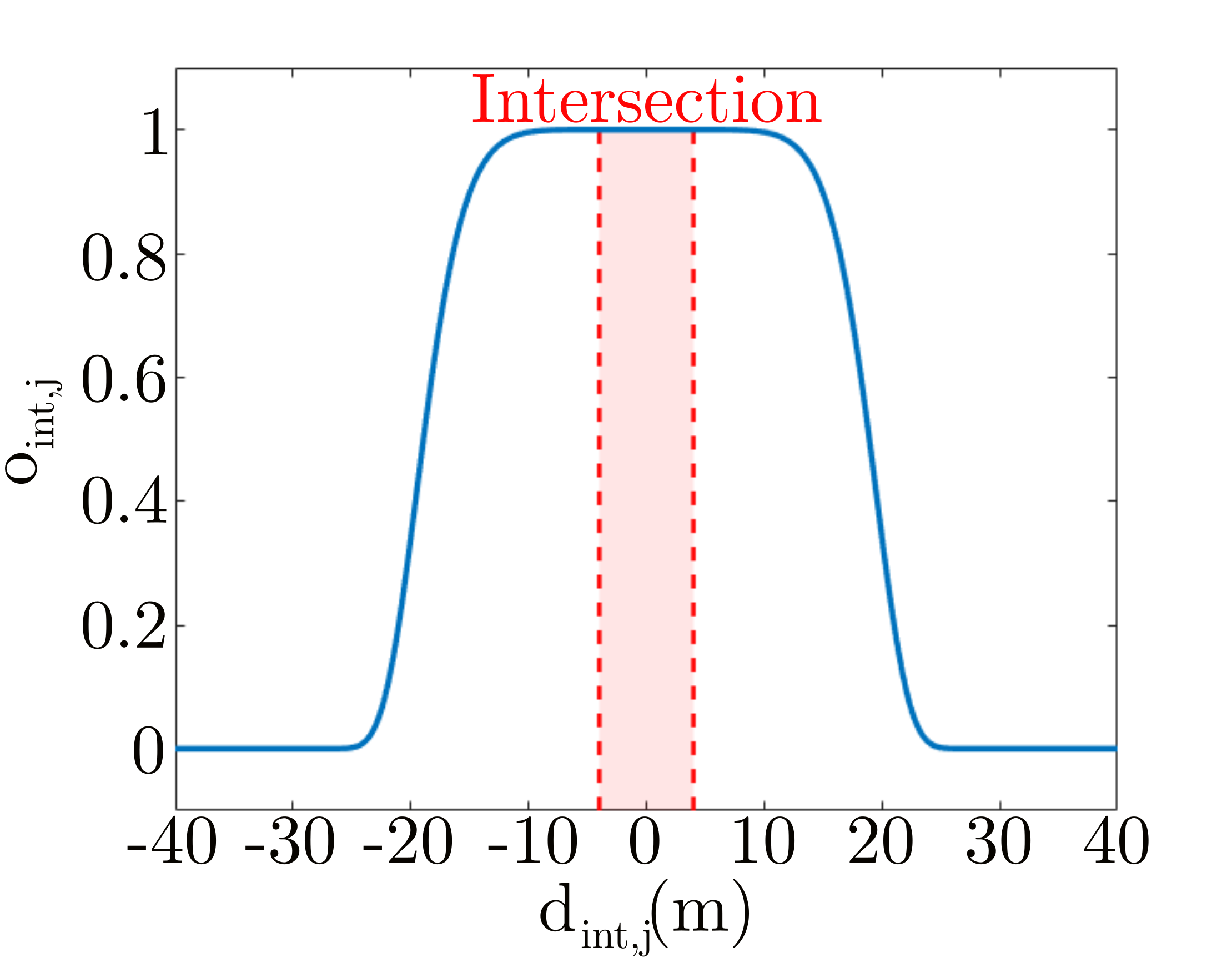}
    \caption{Shape of the per-actor intersection occupancy coefficient \(o_{\mathrm{int},j}\) as a function of signed distance to the intersection center.}
    \label{fig:occupancy_function}
\end{figure}

To map predicted distances \eqref{eq:signed_distance} into a smooth measure of intersection occupancy, at every time instance $k$, each actor $j$ is assigned an occupancy coefficient \(o_{\mathrm{int},j}(k)\in[0,1]\) computed using a generalized Gaussian function (Fig. \ref{fig:occupancy_function}):
\begin{subequations}\label{eq:occ_function}
\begin{alignat}{2}
o_{\mathrm{int},j}(k)
&=
\exp\!\left(
-\left|
\frac{d_{\mathrm{int},j}(k)-d_{\mathrm{shift}}}{\gamma_{\mathrm{occ}}}
\right|^{c_{\mathrm{occ}}}
\right),
&&
\\[0.25em]
d_{\mathrm{spread}}
&= \frac{d_{\mathrm{pre}}+d_{\mathrm{post}}}{2},
\mspace{30mu}d_{\mathrm{shift}}
= \frac{d_{\mathrm{pre}}-d_{\mathrm{post}}}{2},
\\
\gamma_{\mathrm{occ}}
&=
\frac{d_{\mathrm{spread}}}{\left(|\ln(\varepsilon_{\mathrm{occ}})|\right)^{1/c_{\mathrm{occ}}}}.
\end{alignat}
\end{subequations}
where \(c_{\mathrm{occ}}>0\) (typically an even integer) shapes the decay, while the parameters \(d_{\mathrm{pre}}\) and \(d_{\mathrm{post}}\) determine, respectively, how far before and after the intersection center an actor should still influence the conflict zone. Finally, \(\varepsilon_{\mathrm{occ}}\) is a small value defining the tail magnitude at the spread distance. Note that, if \(d_{\mathrm{pre}} \neq d_{\mathrm{post}}\), the resulting metric $o_{int,j}$ is slightly asymmetric and its peak is shifted with respect to the intersection center. This can be used to anticipate the conflict zone depending on the conventions chosen for signed distances.

In the end, we can define a safe global occupancy metric as the maximum over all selected actors
\begin{equation}
    o_{\mathrm{int}}^{*}(k) = \max_j \, o_{\mathrm{int},j}(k).
    \label{eq:global_occ}
\end{equation}
This time-dependent coefficient \(o_{\mathrm{int}}^{*}(k)\) is used to scale the intersection potential over the prediction horizon: when any relevant actor is predicted to be close to the intersection, the coefficient grows and strengthens the repulsive barrier; as the conflict zone clears, the coefficient decreases smoothly, ensuring an efficient optimization problem.
\section{Intersection TV-APF formulation}
\label{sec:int_TV-APF}
\begin{figure}[]
    \centering
    \includegraphics[width=0.7\linewidth]{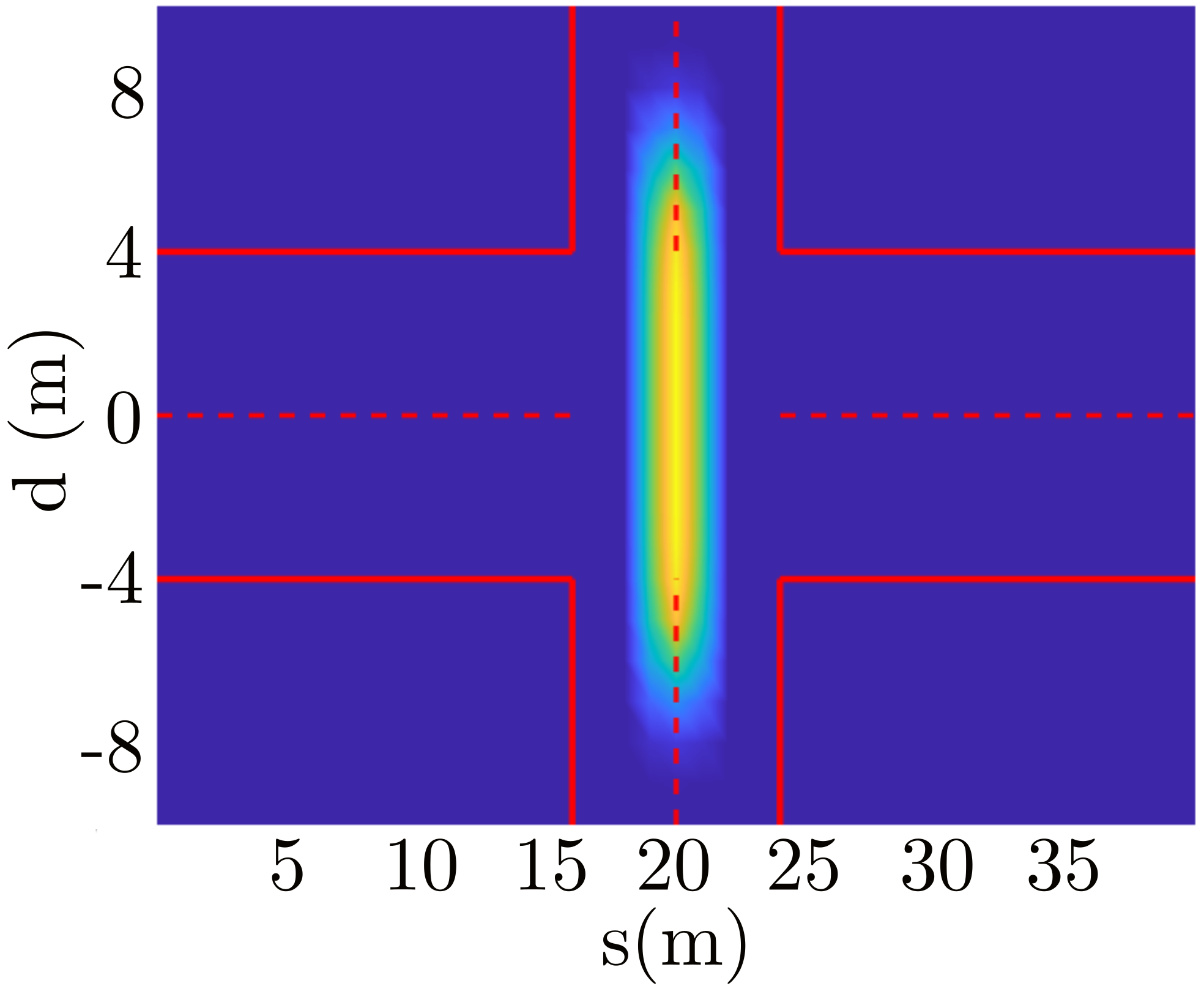}
    \caption{Top view of the intersection TV-APF $W_{int}(k)$ for a specific time instant $k$ when the intersection is fully occupied}
    \label{fig:apf_int_top}
\end{figure}

We model intersection interactions as TV-APF \cite{CostaTVAPF_prev}, consistently with the framework introduced in Section~\ref{sec:prel}. The amplitude of the field is driven by the predicted occupancy coefficient \eqref{eq:global_occ}. From the optimizer perspective, the intersection rule is encoded as a smooth, differentiable penalty that increases near the conflict zone when other vehicles are predicted to occupy it. This design allows behavioral choices, e.g., slowing down vs. proceeding vs. stopping, to emerge from a unified optimization process, avoiding discrete decision switches and maintaining a smooth cost.
Considering \((s,d)\) as the position of the ego vehicle in Frenet coordinates, the intersection TV-APF $W_{\mathrm{int}}$ at prediction step \(k\)  (Fig. \ref{fig:apf_int_top}) is defined as a two-dimensional Gaussian function:
\begin{equation}
    W_{\mathrm{int}}(k) =
    K_{\mathrm{int}}o^{*}_{\mathrm{int}}(k)
    \exp\!\Bigg(
    -\bigg|\frac{s - s_{\mathrm{int}}}{\gamma_{\mathrm{int},s}}\bigg|^{c_{\mathrm{int}}}
    -\bigg|\frac{d}{\gamma_{\mathrm{int},d}}\bigg|^{c_{\mathrm{int}}}
    \Bigg),
    \label{eq:apf_int}
\end{equation}
Where $K_{\mathrm{int}}$ determines the strenght of the overall field, \(c_{\mathrm{int}}\) controls the field shape, instead the decay rates \(\gamma_{\mathrm{int},s}\) and \(\gamma_{\mathrm{int},d}\) are selected from the desired spreads \(\sigma_{\mathrm{int},s}\), \(\sigma_{\mathrm{int},d}\) and tail values \(\varepsilon_{\mathrm{int},s}\), \(\varepsilon_{\mathrm{int},d}\) according to:
\begin{equation}
\gamma_{\mathrm{int},i}
=
\frac{\sigma_{\mathrm{int},i}}
{\left(|\ln\left(\varepsilon_{\mathrm{int},i}/K_{\mathrm{int}}\right)\right|)^{1/c_{\mathrm{int}}}},
\qquad i \in \{s,d\}.
\end{equation}

The occupancy scaling \(o^{*}_{\mathrm{int}}(k)\) makes the potential \textit{time-varying}: when the intersection is predicted to be occupied, the repulsive field strengthens and acts as a virtual barrier; when the conflict zone becomes free, the potential fades out smoothly. The formulation of the intersection field defined in \eqref{eq:apf_int}, well suits within the FHOCP cost formulation in \eqref{eq:cost}.

\section{Simulation Results}
In this section, we report two simulation tests to illustrate the effectiveness of the proposed unified DM-TP formulation for unsignalized intersections\footnote{A video for the bird's eye view of the scenarios is available at https://youtu.be/nYBbVTMabuU}. 
The simulation study is carried out in MATLAB/Simulink, exploiting the hierarchical control architecture introduced in~\cite{CostaTVAPF_prev} using \eqref{eq:nlp_complete} as Local Trajectory Planner (LTP) (with parameters detailed in Table \ref{tab:LTP-FHOCP}) and employing as motion controller a MPC strategy. Traffic scenarios are generated with Automated Driving Toolbox~\cite{MathWorks2026ADT}, while the FHOCP is implemented in CasADi~\cite{Andersson2019} and solved with Ipopt~\cite{Waechter2006}. For simulation, we consider all roads with a lane width of $4$ m and a maximum speed limitation of $12.5$ m/s; in addition, both the ego vehicle and actors are subject to the kinematic limits in Table \ref{tab:limits}.\\
We recall that each actor at intersections and its motion prediction over the horizon $H_P$ is encoded with the $o_{int}$ \eqref{eq:global_occ} coefficient and generates the barrier field $W_{int}$ \eqref{eq:apf_int} within the LTP FHOCP. At each planning cycle $T_P$, updated data on the positions and velocities of the actors are used to refine their prediction and the corresponding occupancy signal $o_{int}$.

The simulation test consists of two different intersections that are joined together by a straight road, which is not described here.
\begin{table}[]
    \vspace{5pt}
    \centering
    \caption{LTP (FHOCP) parameters}
    \label{tab:LTP-FHOCP}
    \begin{tabular}{l c c c}
        \hline
        Parameter & Unit & Value \\
        \hline
        Prediction horizon $H_P$  & \(\mathrm{s}\)  &  35 \\
        Sample time $T_s$  & \(\mathrm{s}\)  &  0.5\ \\
        Planner period $T_P$  & \(\mathrm{s}\)  & 5\ \\
        \hline
    \end{tabular}
\end{table}
\begin{table}[]
    \centering
    \caption{Vehicles kinematic limits}
    \label{tab:limits}
    \begin{tabular}{l c c c}
        \hline
        Parameter & Unit & Minimum & Maximum \\
        \hline
        Speed  & \(\mathrm{m/s}\)  & 0 & 12.5   \\
        Longitudinal acc. & \(\mathrm{m/s^2}\)  & -0.9 & 0.9\ \\
        Jerk & \(\mathrm{m/s^3}\)  & -0.9 & 0.9\ \\
        Yaw rate  & \(\mathrm{deg/s}\)& -4.44 & 4.44\\
        Steering angle & \(\mathrm{deg}\) & -24.5 & 24.5\\
        \hline
    \end{tabular}
\end{table}
\begin{figure}[t]
    \centering
    \includegraphics[width=1\linewidth]{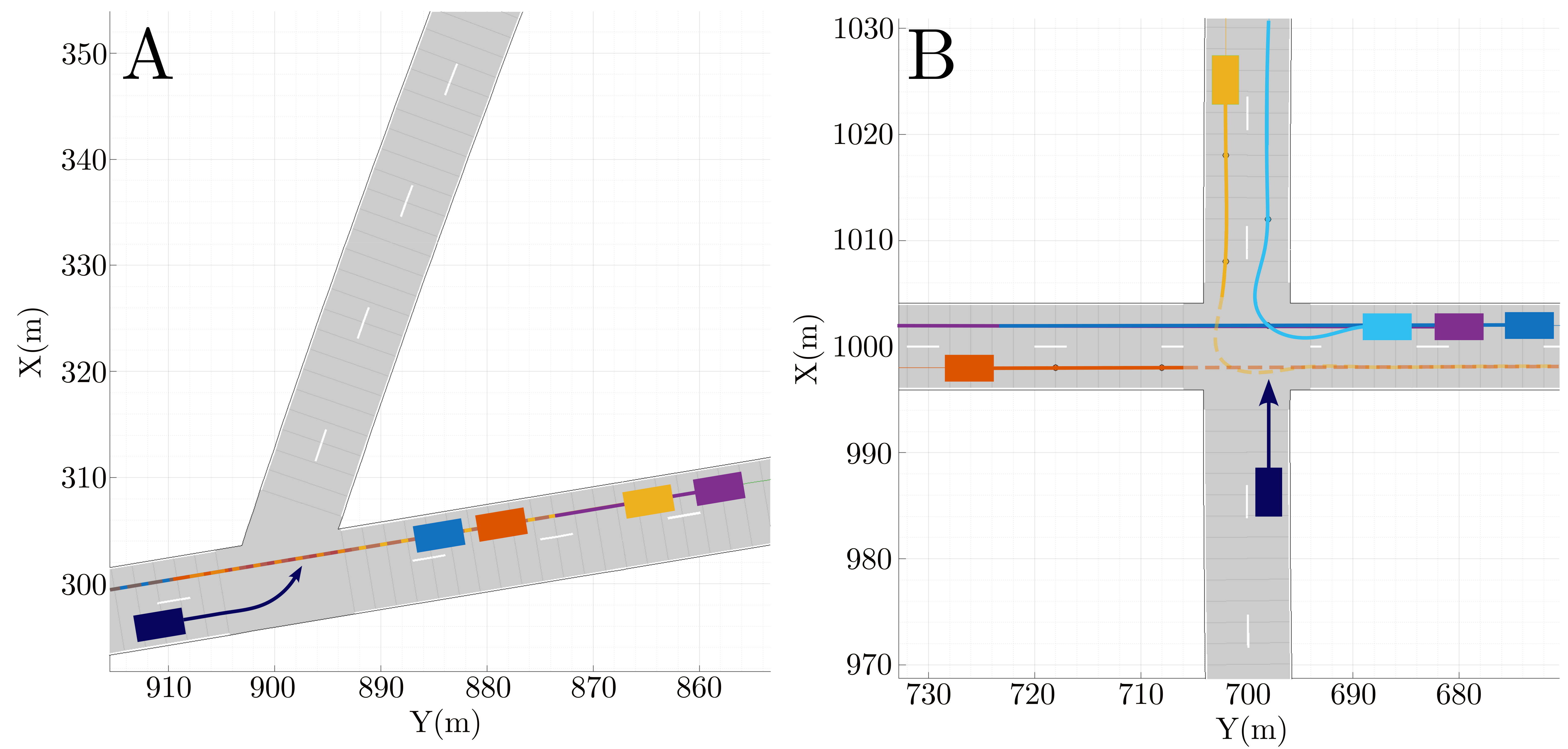}
    \caption{Simulation snapshots of two unsignalized-intersection scenarios: (A) The ego vehicle (dark blue) performs a left turn by exploiting an available traffic gap. (B) The ego vehicle crosses a congested four-way intersection while handling interacting traffic according to right-of-way rules. The arrow indicates the ego intended maneuver. For the surrounding actors, trajectories are shown with the same color as the corresponding vehicle; dotted trajectories indicate vehicles that stop to yield to the ego and then proceed only after the ego has left the conflict zone}
    \label{fig:scenario-overview}
\end{figure}
\begin{figure}
    \centering
    \includegraphics[width=\linewidth]{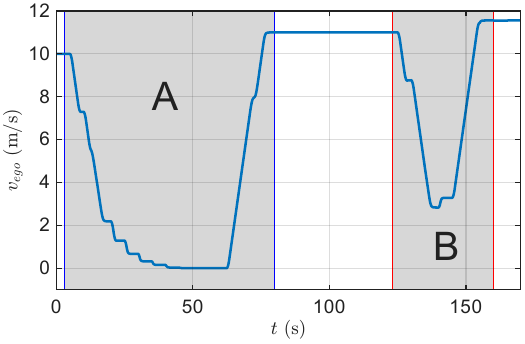}
    \caption{Time evolution of the ego speed in the simulated scenario. The speed profile highlights the behavior modulation emerging from the FHOCP for scenarios A and B.
    Specifically, it shows (A) a stop-and-go behavior at the first intersection to exploit a safe traffic gap, followed by (B) a second speed reduction at the four-way intersection to account for right-of-way interactions.
    }
    \label{fig:simulation-speed}
\end{figure}

    \emph{A) Left Turn Using Traffic Gap} We first consider the three-way irregular intersection in Fig. \ref{fig:scenario-overview} A, in which the ego vehicle must make a left turn on a secondary road.
    Traffic is placed on the main road with a variable gap between the actors. The ego vehicle must turn to the left as soon as it is safe to do so (i.e., when a sufficiently large gap in the traffic appears).
    Due to the effect of $W_{int}$, the ego vehicle speed $v_{ego}$ progressively slows down from its reference of $10$ m/s to $0$ m/s, as shown in Fig. \ref{fig:simulation-speed}. Finally, at $t=60$ s, the updated prediction ensures a sufficient gap, and the ego vehicle promptly accelerates until its reference speed of $11$ m/s.

    \emph{B) Congested Intersection} At time $t=125$ s, the ego vehicle reaches the four-way unsignalized intersection in Fig. \ref{fig:scenario-overview} B. This time, the ego vehicle must yield to three vehicles coming from the right (in cyan, purple, and blue) and has precedence over two vehicles coming from the other two branches (in orange and yellow). According to the rules in Section \ref{sec:int_actor_selection}, even if the ego vehicle has precedence over the orange and yellow actors, they are considered \emph{potential interferers} until they are close to stopping. The resulting behavior, generated by the FHOCP, is to decelerate progressively.
    As soon as the updated data show that the three actors on the right have passed and that the other two actors are approaching a full stop, at time $t=140$ s, the ego vehicle resumes its speed tracking to the reference speed of $11.5$ m/s, successfully crossing the intersection without risk.

\section{Conclusion}

This paper presented a unified DM and TP framework for automated vehicles at unsignalized intersections based on a FHOCP augmented with an intersection-oriented TV-APF. The proposed formulation combines intersection-aware actor selection, short-horizon occupancy estimation, and a smooth conflict-zone potential to encode right-of-way interactions directly within the FHOCP cost. In this way, behaviors such as slowing down, yielding, stopping, and proceeding through the intersection emerge from a single optimization problem, without requiring discrete decision switching. The reported simulations show that the proposed approach is capable of effectively handling different types of intersection and traffic conditions while keeping the ego vehicle safe.

\bibliographystyle{IEEEtran}
\bibliography{IEEEabrv,bibliography}

\begin{thebibliography}{10}
\providecommand{\url}[1]{#1}
\csname url@rmstyle\endcsname
\providecommand{\newblock}{\relax}
\providecommand{\bibinfo}[2]{#2}
\providecommand\BIBentrySTDinterwordspacing{\spaceskip=0pt\relax}
\providecommand\BIBentryALTinterwordstretchfactor{4}
\providecommand\BIBentryALTinterwordspacing{\spaceskip=\fontdimen2\font plus
\BIBentryALTinterwordstretchfactor\fontdimen3\font minus \fontdimen4\font\relax}
\providecommand\BIBforeignlanguage[2]{{%
\expandafter\ifx\csname l@#1\endcsname\relax
\typeout{** WARNING: IEEEtran.bst: No hyphenation pattern has been}%
\typeout{** loaded for the language `#1'. Using the pattern for}%
\typeout{** the default language instead.}%
\else
\language=\csname l@#1\endcsname
\fi
#2}}

\bibitem{AlSharman2024}
M.~Al-Sharman, L.~Edes, B.~Sun, V.~Jayakumar, H.~Tahir, M.~A. Daoud, B.~J. Emran, D.~Rayside, and W.~Melek, ``Autonomous driving at unsignalized intersections: A review of decision-making challenges and reinforcement learning-based solutions,'' \emph{IEEE Transactions on Automation Science and Engineering}, vol.~23, pp. 1434--1461, 2026.

\bibitem{ChenHu2024}
S.~Chen, X.~Hu, J.~Zhao, R.~Wang, and M.~Qiao, ``A review of decision-making and planning for autonomous vehicles in intersection environments,'' \emph{World Electric Vehicle Journal}, vol.~15, no.~3, p.~99, 2024.

\bibitem{RiosTorresMalikopoulos2017}
J.~Rios-Torres and A.~A. Malikopoulos, ``A survey on the coordination of connected and automated vehicles at intersections and merging at highway on-ramps,'' \emph{IEEE Transactions on Intelligent Transportation Systems}, vol.~18, no.~5, pp. 1066--1077, 2017.

\bibitem{ZhaoMACPO2024}
R.~Zhao, Y.~Li, F.~Gao, Z.~Gao, and T.~Zhang, ``Multi-agent constrained policy optimization for conflict-free management of connected autonomous vehicles at unsignalized intersections,'' \emph{IEEE Transactions on Intelligent Transportation Systems}, vol.~25, no.~6, pp. 5374--5388, 2024.

\bibitem{TianGameTheory2022}
R.~Tian, N.~Li, I.~Kolmanovsky, Y.~Yildiz, and A.~R. Girard, ``Game-theoretic modeling of traffic in unsignalized intersection network for autonomous vehicle control verification and validation,'' \emph{IEEE Transactions on Intelligent Transportation Systems}, vol.~23, no.~3, pp. 2211--2226, 2022.

\bibitem{LiPeng2024}
S.~Li, K.~Peng, F.~Hui, Z.~Li, C.~Wei, and W.~Wang, ``A decision-making approach for complex unsignalized intersection by deep reinforcement learning,'' \emph{IEEE Transactions on Vehicular Technology}, vol.~73, no.~11, pp. 16\,134--16\,147, 2024.

\bibitem{ChenXu2024}
X.~Chen, B.~Xu, M.~Hu, Y.~Bian, Y.~Li, and X.~Xu, ``Safe efficient policy optimization algorithm for unsignalized intersection navigation,'' \emph{IEEE/CAA Journal of Automatica Sinica}, vol.~11, no.~9, pp. 2011--2026, 2024.

\bibitem{TangUncertainty2023}
X.~Tang, G.~Zhong, S.~Li, K.~Yang, K.~Shu, D.~Cao, and X.~Lin, ``Uncertainty-aware decision-making for autonomous driving at uncontrolled intersections,'' \emph{IEEE Transactions on Intelligent Transportation Systems}, vol.~24, no.~9, pp. 9725--9735, 2023.

\bibitem{MPCsw}
X.-F. Wang, W.-H. Chen, J.~Jiang, and Y.~Yan, ``High-level decision-making for autonomous overtaking: An mpc-based switching control approach,'' \emph{IET Intelligent Transport Systems}, vol.~18, no.~7, pp. 1259--1271, 2024.

\bibitem{APF0}
O.~Khatib, ``Real-time obstacle avoidance for manipulators and mobile robots,'' \emph{The International Journal of Robotics Research}, vol.~5, no.~1, pp. 90--98, 1986.

\bibitem{Zhang2025DAWM}
X.~Zhang, Z.~Wu, H.~Hu, J.~Yang, and P.~Wang, ``Integrating driving-aware world model with {MPC} for autonomous driving at unsignalized {T}-intersections,'' \emph{IEEE Transactions on Intelligent Transportation Systems}, vol.~26, no.~12, pp. 22\,219--22\,231, 2025.

\bibitem{JeongYi2021}
Y.~Jeong and K.~Yi, ``Target vehicle motion prediction-based motion planning framework for autonomous driving in uncontrolled intersections,'' \emph{IEEE Transactions on Intelligent Transportation Systems}, vol.~22, no.~1, pp. 168--177, 2021.

\bibitem{Raja2024}
G.~Raja, K.~Raja, M.~R. Kanagarathinam, J.~Needhidevan, and P.~Vasudevan, ``Advanced decision making and motion planning framework for autonomous navigation in unsignalized intersections,'' \emph{IEEE Access}, vol.~12, pp. 158\,657--158\,668, 2024.

\bibitem{CostaTVAPF_prev}
\BIBentryALTinterwordspacing
D.~Costa, F.~Cerrito, M.~Canale, and C.~Novara, ``Unifying decision making and trajectory planning in automated driving through time-varying potential fields,'' 2026. [Online]. Available: \url{https://arxiv.org/abs/2603.13136}
\BIBentrySTDinterwordspacing

\bibitem{MathWorks2026ADT}
{The MathWorks, Inc.}, ``{Automated Driving Toolbox},'' \url{https://www.mathworks.com/products/automated-driving.html}, 2026, accessed: 2026-03-12.

\bibitem{Andersson2019}
J.~A.~E. Andersson, J.~Gillis, G.~Horn, J.~B. Rawlings, and M.~Diehl, ``{CasADi}: A software framework for nonlinear optimization and optimal control,'' \emph{Mathematical Programming Computation}, vol.~11, no.~1, pp. 1--36, 2019.

\bibitem{Waechter2006}
A.~W{\"a}chter and L.~T. Biegler, ``On the implementation of an interior-point filter line-search algorithm for large-scale nonlinear programming,'' \emph{Mathematical Programming}, vol. 106, no.~1, pp. 25--57, 2006.

\end{thebibliography}

\end{document}